 \documentclass[final,5p,times,twocolumn,authoryear]{elsarticle}

\usepackage{amssymb}
\usepackage{amsmath}
\usepackage{epstopdf}
\usepackage{etoolbox}
\usepackage{natbib}
\usepackage{hyperref,url}
\usepackage{xcolor}
\newcounter{bibcount}
\makeatletter
\patchcmd{\@lbibitem}{\item[}{\item[\hfil\stepcounter{bibcount}{\thebibcount.}}{}{}
\renewcommand\NAT@bibsetup%
   [1]{\setlength{\leftmargin}{\bibhang}\setlength{\itemindent}{-\parindent}%
      \setlength{\itemsep}{\bibsep}\setlength{\parsep}{\z@}}
\makeatother
\journal{Physics Letter B}

\begin{document}

\begin{frontmatter}



\title{Graviton Condensate Stars and Its Gravitational Echoes}


\author{Muhammad Fitrah Alfian Rangga Sakti} 
\ead{fitrahalfian@gmail.com}



\address{High Energy Physics Theory Group, Department of Physics, Faculty of Science, Chulalongkorn University, Bangkok 10330, Thailand}
\address{Department of Physics and Astronomy, University of Waterloo, Waterloo, Ontario, N2L 3G1, Canada}
\address{Perimeter Institute for Theoretical Physics, Waterloo, Ontario, N2L 2Y5, Canada}

\begin{abstract}
We construct the exact stellar configurations that contain an ordinary perfect-fluid matter that interacts minimally with a condensate of gravitons with distinct pressure conditions on the surface. We propose vanishing transverse pressure on the surface for, namely graviton condensate type 1 and vanishing radial pressure on the surface for type 2. The condition for the radial pressure of type 1 requires the existence of a thin shell that will balance the pressure discontinuity while for type 2, the discontinuity on transverse pressure does not require the additional thin shell. It is found that the Buchdahl inequality of the resulting stellar configurations depends on the parameter related to the graviton condensate, such that we can find the ultra-compact regime of the stellar models. Moreover, the echo time and echo frequency within the ultra-compact regime are computed. At the same compactness, it is found that the presence of the graviton condensate will delay the gravitational echoes for type 2 and will expedite the gravitational echoes for type 1 compared to constant density star, $\tau_{echo2}>\tau_{CDS}>\tau_{echo1}$. Furthermore, the gravitational perturbation of a massless scalar wave is also investigated to support these results. These results could open more opportunities for the observational study of graviton in the near future, mostly from the compact astrophysical objects.

\end{abstract}


\end{frontmatter}


\section{Introduction}
\label{sec:intro}

It was originally proposed by Dvali, Gomez, and their colleagues in several intriguing articles that black holes (BHs) could be considered as a condensate of gravitons at the critical point of a quantum phase transition \citep{DvaliGomezPhysLettB2012,DvaliGomezPhysLett2013,DvaliGomezFortschPhys2013,DvaliGomezPRD2013,DvaliGomezEPJC2014}. It is very enticing idea which could lead to new understanding of some compact objects such as black holes (BHs). Due to this new understanding, the Hawking radiation of black holes can be argued as the radiation of the gravitons from the condensate. Since one can consider the non-vanishing Hawking radiation from this model, the Bekenstein-Hawking entropy, generally the thermodynamics of BHs, can also be explained using the method from condensed matter physics. The gravitons contained in the condensate are basically deemed as off-shell gravitons which do not satisfy the classical energy-momentum relation ($P^2\neq 0$).

In this novel point of view,  BH is assumed to be at the critical point of the quantum phase transition. BHs are grasped in term of number of off-shell gravitons with the number of $ N $ presented in the Bose-Einstein condensate (BEC). The number of gravitons is given as $ N \sim M r_g/\hbar $ where $ r_g $ is the gravitational radius and $M$ is related to the mass of the BHs. The interaction between gravitons is measured by the coupling constant $ \alpha \sim L_P^2/\lambda^2 $ where $ L_P^2=\hbar G ~[c\equiv 1] $ (Planck length) and $ \lambda $ is the characteristic wavelength of the gravitons. The wavelength of the gravitons is also related to the number of the gravitons given as $\lambda \sim \sqrt{N} L_P$. Therefore, based on this theory, BHs must always satisfy $ N \alpha \sim 1 $ as the critical condition for maximal $N$. Moreover, it can't enter into strong coupling regime $(N\alpha\gg 1)$. For large wavelengths, gravitons act as if they were free quasiparticles. Furthermore, in \cite{AlfaroEspriuCQG2018,AlfaroEspriuIJMPA2020}, they have used this understanding to construct a geometrical model of the graviton condensate. They have contended that the resulting equation of motion derived from the action is interpreted as graviton condensate trapped in by the BHs' gravitational field. They also have formulated the thermodynamics of such geometrical model and found that the Bekenstein-Hawking entropy, BH mass, and Hawking temperature are given by
\begin{equation}
	S_{BH} \sim N, ~~~ M\sim N^{1/2} , ~~~  T_H \sim N^{-1/2},\label{eq:SMT}\
\end{equation}
respectively, for a Schwarzschild BH with a condensate of gravitons \citep{AlfaroMancillaEPJC2021}. The Hawking temperature is considered as the leakage of the gravitons from the condensate. Moreover, they have succeeded in associating the volume and pressure with the graviton condensate using the horizon thermodynamic approach \citep{PadmanabhanCQG2002,PadmanabhanMPLA2002}.

In the gravitational wave physics era, the characterization of the gravitational wave signal emitted by compact binary sources plays a important role in understanding the compact star properties. Generally, there are three important regimes related to the compactness $C$ of the compact objects which are the black hole regime, regime between black hole and the Buchdahl limit, and the regime between the Buchdahl limit and photon sphere \citep{MannarelliPRD2018,BoraMNRAS2021,AlhoPRD2022}. Gravitational waves' echoes may exist on the stars within the ultra-compact regime which is the regime within the Buchdahl limit and photon sphere. The photon sphere is the unstable circular null geodesic of the exterior BH spacetime metric. The gravitational wave signal will be emitted in the final state of a compact binary merger. The post-merger ringdown waveform is believed to be initially identical to that of a BH with the modifications encoded in subsequent echoes when the resulting object from the merger process is sufficiently compact to have a photon sphere. The simplest model of the ultra-compact objects to produce the gravitational waves echoes is the constant density star (CDS) \citep{UrbanoVeermaeJCAP2019}. The regime for the production of the echoes is given in the range $1/3 \leq C \leq 4/9$, denoting the photon sphere and Buchdahl limit \citep{MannarelliPRD2018,UrbanoVeermaeJCAP2019}.

Within this work, we employ the geometrical model of the graviton condensate proposed in \cite{AlfaroEspriuCQG2018} in the stellar configurations. We assume that the matter contribution comes from the graviton condensate and the perfect-fluid matter with constant energy density. We want to compare the results to that of vanishing graviton condensate case, i.e. CDS. We consider two different stellar configurations based on the pressure conditions on the surface of the stars. The first condition that is imposed is the vanishing transverse pressure on the boundary while another one is the vanishing radial pressure on the boundary. The Buchdahl inequalities of these stellar configurations are also investigated. Furthermore, in the ultra-compact regime under the Buchdahl bound, we examine the gravitational-echo time and frequency to see the effect of the graviton condensate. The gravitational perturbation of a massless scalar wave equation on the stellar configurations is also investigated to support more of the results on the echoes' production.

\section{Graviton Condensate Stars}
In this section, we will derive a stellar configuration which contains a perfect-fluid matter and the graviton condensate. We will start from the geometrical model of graviton condensate which we will adopt. The graviton condensate model is introduced originally as a geometrical model in \cite{AlfaroEspriuCQG2018} as a perturbation on the spacetime metric to produce Schwarzschild BH which contains graviton condensate. Although this approximation is actually deviated from the genuine concept of Dvali and Gomez, this geometrical approximation is proposed to obtain novel geometrical solution containing graviton condensate. In this geometrical model, the graviton condensate possesses its own Lagrangian coming from the spacetime perturbation which leads to the a specific energy-momentum tensor.

In this regard, the metric is split into two parts as $
g_{\mu\nu} = \bar{g}_{\mu\nu} + h_{\mu\nu} $ where $ \bar{g}_{\mu\nu} $ is the background metric and $ h_{\mu\nu} $ is the quantum fluctuation indicating the graviton condensate. In order to raise or lower the indices, one needs to employ the total metric $g_{\mu\nu}$. In terms of action, the graviton condensate can be described as follows
\begin{equation}
	S_G = -\frac{1}{8}\int d^4 x \sqrt{-\bar{g}} \mu(x) h_{\alpha \beta}h^{\alpha \beta}. \label{eq:gravitonaction}
\end{equation}
$ \mu(x) $ is just a scalar function that, at first, was interpreted as chemical potential in \cite{AlfaroEspriuIJMPA2020}. Nevertheless, it is pointed out that $\mu$ does not have to be interpreted as chemical potential because the actual chemical potential vanishes at the critical point of the quantum phase transition. We want to emphasize that the gravitons contained in the condensate are considered as the off-shell gravitons which do not satisfy the classical energy-momentum relation.

Using the action provided in (\ref{eq:gravitonaction}) plus the usual Einstein-Hilbert action, one can obtain the following equation of motion \citep{AlfaroEspriuCQG2018}
\begin{equation}
	G_{\alpha\beta}(g) = \Sigma~\mu(x)(h_{\alpha\beta} -h_{\alpha\sigma}h^\sigma_\beta), \label{eq:EoMgraviton}
\end{equation}
where $ G_{\alpha\beta} = R_{\alpha\beta} - \frac{1}{2} R g_{\alpha\beta} $ and $ \Sigma \equiv \sqrt{\frac{-\bar{g}}{-g}} $. The \textit{rhs} of Eq. (\ref{eq:EoMgraviton}) is produced from the metric fluctuation that can be described as the effective matter tensor from the graviton condensate. The equation of motion (\ref{eq:EoMgraviton}) is interpreted as the Gross-Pitaevskii equation to portray the properties of BH which contains graviton condensate. Besides Schwarzschild solution, in \cite{AlfaroEspriuIJMPA2020}, they have extended the solution with non-vanishing electric charge and cosmological constant which contain graviton condensate. To be specific, the fluctuation of the metric possesses only non-vanishing time-like and radial parts as $h^t_t = h^r_r = b $  where $ b $ is a constant related to the graviton condensate, taken between 0 to 1. Generally, one can write the effective matter tensor as \citep{AlfaroEspriuCQG2018}
\begin{equation}
	T_\beta^\alpha = \left(-\frac{b}{\kappa r^2},-\frac{b}{\kappa r^2},0,0 \right). \label{eq:EMfluctuation}
\end{equation}
We consider the metric convention ($-,+,+,+$) and $G=c=1$ for the rest of the work. As we have mentioned in the beginning of this section, we further consider a perfect-fluid matter in the matter contribution. The simplest matter tensor of the perfect fluid is given as follows
\begin{equation}
	T^{\mu}_{\nu} = p \delta^{\mu}_{\nu} + \left(\rho + p \right)u^\mu u_\nu , \label{eq:perfectfluid}
\end{equation}
where $ \rho $ and $ p $ are the energy density and pressure, respectively.

We assume a spherically symmetric space-time as the interior which reads as
\begin{equation}
	ds^2 = -e^{2\nu(r)}dt^2 + e^{2\lambda(r)}dr^2+r^2d\theta^2 +r^2 \sin^2\theta d\phi^2.\label{eq:generalmetric}\
\end{equation}
By employing Eqs. (\ref{eq:EMfluctuation}), (\ref{eq:perfectfluid}), and (\ref{eq:generalmetric}) into the field equation (\ref{eq:EoMgraviton}), we can obtain the following set of equations
\begin{eqnarray}
	e^{-2\lambda}\left({\frac{2}{r}\lambda'-\frac{1}{r^2}}\right)+\frac{1}{r^2}&=&\kappa\rho+\frac{b}{r^2},\label{eq:Einstein1} \\ 
	e^{-2\lambda}\left({\frac{2}{r}v'+\frac{1}{r^2}}\right)-\frac{1}{r^2}&=&\kappa p -\frac{b}{r^2},\label{eq:Einstein2} \\
	e^{-2\lambda}\left({v''+{v'}^2-v' \lambda'+\frac{v'}{r}-\frac{\lambda'}{r}}\right)&=&\kappa p, \label{eq:Einstein3}
\end{eqnarray}
where $\kappa=8\pi$. To solve the given set of non-linear equations (\ref{eq:Einstein1})-(\ref{eq:Einstein3}), we  generally impose the uncompressible energy density for the perfect-fluid matter, ($\rho =$ constant) and different conditions of the total pressure on the surface. It is forefront to note that the exterior solution we consider here is the Schwarzschild BH because we want that the graviton condensate is confined within the stars' interior only. On the other hand, the exterior solution is not a black hole with graviton condensate as given in \cite{AlfaroEspriuIJMPA2020,AlfaroMancillaEPJC2021}.

We choose the constant energy density for the perfect-fluid matter is because we want to obtain the simplest analytical star model which contains the graviton condensate. CDS is the simplest analytical star model which contains only perfect-fluid matter, but it is rich of physical applications behind it, for example it has been employed to study gravitational echoes \citep{UrbanoVeermaeJCAP2019,PaniFerrariCQG2018} and gravastar by assuming the compactness to be one \citep{MazurMottolaPNAS2004,MazurMottolaCQG2015,MazurMottolaUniv2023}. By adding the graviton condensate to the interior solution, we believe that this will benefit the study of graviton in terms of stellar configuration.

It is worth noting that since the energy density from the graviton condensate is dependent on radial coordinate as given in Eq. (4), therefore, the total energy density for this graviton condensate star is still dependent on the radial coordinate, mainly coming from the graviton condensate contribution. In this case, one can still have the total energy density to be maximal at the core and decrease along the way to the surface of the star. So, the total energy density is not uniform across the entire interior region. Furthermore, the uncompressible energy density for the perfect-fluid matter has also been used in dark energy star \citep{YazadjievPRD2011}.

The first condition we propose is the vanishing transverse pressure on the surface, $ p_{t}(R)=0 $. In this first configuration, we obtain the following graviton condensate star type 1,
\begin{eqnarray}
	ds^2_1 &=& -\left[\frac{3}{2}f(R)-\frac{1}{2}f(r) \right]^2  dt^2  + \frac{dr^2}{1-b- Ar^2 } \nonumber\\
	& +& r^2\left(d\theta ^2 +\sin^2\theta d\phi^2 \right), \label{eq:gavitonstargmetric1}\
\end{eqnarray}
where $f(r)=\sqrt{1-b-Ar^2}$ and $f(R)=f(r)|_{r=R}=\sqrt{1-b-AR^2}$. The total energy density, radial pressure and transverse pressure of the star are given by
\begin{eqnarray}
	\rho_{total}(r) = \rho+\frac{b}{\kappa r^2},~	p_{r1} (r)= p_1(r) -\frac{b}{\kappa r^2}, ~	p_{t1}(r) = p_1(r) ,\ \label{eq:pressgrav1}
\end{eqnarray}
where
\begin{eqnarray}
	\rho = \frac{6m}{\kappa R^3} - \frac{3b}{\kappa R^2}, ~~~	p_1(r) = \rho\left[ \frac{f(r)-f(R)}{3f(R)-f(r)} \right],\label{eq:pressmatter1}\
\end{eqnarray}
which correspond with the energy density and pressure of the perfect-fluid matter, respectively. Since the energy density and pressure of the perfect-fluid matter contain the graviton condensate, this implies that the graviton condensate backreacts to the perfect fluid. 
We  define $A=\frac{\kappa\rho}{3}=\frac{2m}{R^3}-\frac{b}{R^2}$. Note that $ m $ is mass parameter and $ R $ is star radius. As we can see that when $b=0$, the line element of the graviton condensate star type 1 reduces to CDS. On the surface of the star, we can obtain the following conditions,
\begin{equation}
	\rho_{total}(R) = \frac{6m}{8\pi R^3}-\frac{2b}{8\pi R^2}, ~~~ p_{r1}(R)=-\frac{b}{8\pi R^2}, ~~~ p_{t1}(R)=0.\label{eq:surfacecon1}
\end{equation}

The second condition we propose is the vanishing radial pressure on the surface, $ p_{r}(R)=0 $. In this configuration, we obtain the following line element for graviton condensate star type 2,
\begin{eqnarray}
	ds^2_{2} &=& -\left[\frac{3 +\frac{b}{AR^2}}{2}f(R) - \frac{1 +\frac{b}{AR^2}}{2}f(r) \right]^2  dt^2  \nonumber\\
	&+& \frac{dr^2}{1-b- Ar^2 }  + r^2\left(d\theta ^2 +\sin^2\theta d\phi^2 \right). \label{eq:gavitonstargmetric2}\
\end{eqnarray}
The expression of the total energy density and pressure of the graviton condensate star type 2 is given by the similar form as Eq. (\ref{eq:pressgrav1}), yet with $1\rightarrow2$. The related energy density of the perfect-fluid matter is similar with Eq. (\ref{eq:pressmatter1}), yet with the following pressure
\begin{equation}
	p_2(r)=\rho\left[ \frac{(1 +\frac{b}{AR^2})f(r)-(1 +\frac{b}{3AR^2})f(R)}{(3 +\frac{b}{AR^2})f(R)-(1 +\frac{b}{AR^2})f(r)} \right],\ \label{eq:pressmatter2}
\end{equation}
Similarly with the graviton condensate star type 1, when $b=0$, the line element of the graviton condensate star type 2 reduces to CDS. On the surface of the star, we can obtain the following conditions,
\begin{equation}
	\rho_{total}(R) = \frac{6m}{8\pi R^3}-\frac{2b}{8\pi R^2}, ~~~ p_{r2}(R)=0, ~~~ p_{t2}(R)=\frac{b}{8\pi R^2}.\label{eq:surfacecon2}
\end{equation}

\begin{figure*}
	\centering
	\begin{tabular}{c c}
		\includegraphics[width=.45\linewidth]{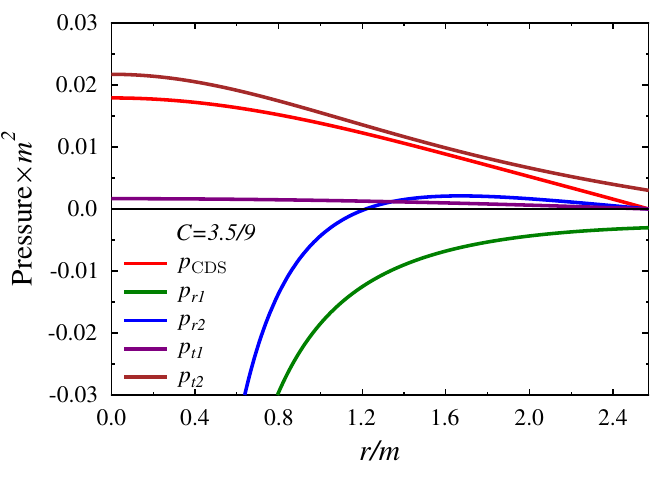} & \includegraphics[width=.45\linewidth]{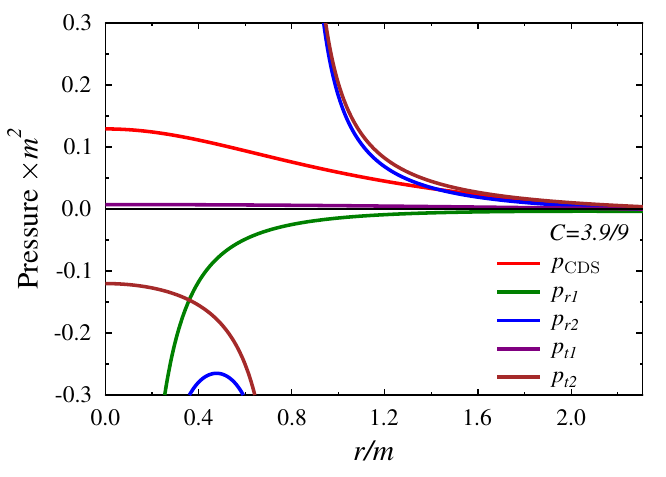}\\
		(a) & (b) \
	\end{tabular}
	\caption{Plot of pressures as a function of $r$ for (a) $C=3.5/9$ and (b) $C=3.9/9$.}
	\label{fig:Pressure}
\end{figure*}
The plot of the pressures of CDS, graviton condensate stars type 1 and type 2 is given in Fig. \ref{fig:Pressure}. We set $b=0.5$ and $C=3.5/9$ for figure (a) and $C=3.9/9$ for figure (b). We have defined the compactness as $C=m/R$. Due to the existence of the graviton condensate, $p_{r1}$ is negative while $p_{r2}$ is negative only from the center of the star to the arbitrary radii. Furthermore, $p_{t1}$ is always positive while $p_{t2}$ becomes negative from the center of the star to an arbitrary radius for high compactness ($C=3.9/9$). The negative pressure also occurs in the compact stars containing dark-energy matter \cite{YazadjievPRD2011,BertolamiPRD2005,LoboCQG2006,DasDebnathEPJC2024}. The negative pressure may imply the violation of the strong-energy condition \cite{HorvatCQG2013,BharRahamanEPJC2015,TudeshkiBordbarPLB2022,SaktiSulaksonoPRD2021}.

Moreover, although the energy density of the perfect fluid is constant, it is obvious that the graviton condensate stars type 1 and type 2 are compressible because $ \rho_{total}=\rho_{total} (r) $. Both stars also possess similar energy density. Because of this, both stars must have similar mass function. The mass function of the stars can be computed by  integrating $ \rho_{total} $ inside the star 
\begin{equation}
	dM = \int^r_0 4\pi r'^2 \rho_{total}dr', 
\end{equation}
which will produce
\begin{equation}
	M(r) = m \frac{r^3}{R^3}+\frac{b r}{2}\left(1- \frac{r^2}{R^2}\right). \label{eq:massfunction}\
\end{equation}
Hence, the total mass at the surface is $M(R)=M=m$. Recall that the exterior solution for both stellar configurations is Schwarzschild BH. It is different with the Schwarzschild BH containing graviton condensate given as follows \citep{AlfaroMancillaEPJC2021}
\begin{eqnarray}
	ds^2 &=& -\left(1-b -\frac{2m}{r}\right)  dt^2  +\left(1-b -\frac{2m}{r}\right)^{-1}dr^2  \nonumber\\
	&+& r^2\left(d\theta ^2 +\sin^2\theta d\phi^2 \right). \label{eq:exteriormetric}\
\end{eqnarray}
For BH solution (\ref{eq:exteriormetric}), the parameter $b$ does not vanish at infinity. We cannot assume $b=1$ for this exterior solution since it will change the sign of $g_{tt}$ and $g_{rr}$. So, for our stellar configurations, it is more suitable to assume the Schwarzschild BH as the exterior solution of the stars for which $b=1$ is still appropriate.

\section{Junction Conditions}
\label{sec:junction}
Both graviton condensate stars possess continuous metric coefficients at the surface. However, as we observe from the pressures on Eq. (\ref{eq:surfacecon1}) for type 1 and on Eq. (\ref{eq:surfacecon2}) for type II, the radial pressure is not continuous for type 1 and transverse pressure is not continuous for type 2. Such discontinuity can be solved by considering the Darmois-Israel formalism \citep{IsraelNuovo1966,IsraelNuovo1967,BerezinPRD1987} to determine the surface stresses at the junction boundary. Hence, the thin shell that separates interior and exterior solutions is considered to exist such that it solves the discontinuity, if the discontinuity occurs. The instrinsic surface stress energy tensor $S_{ij}$ is given by the Lancozs equation,
\begin{equation}
	S^i_j = -\frac{1}{\kappa}\left(K^i_j -\delta^i_j K\right),\label{eq:Lancozs}
\end{equation}
where $K_{ij} = K^+_{ij}-K^-_{ij}$ gives the discontinuous surfaces in the extrinsic curvatures. The signs ``$+$'' and ``$-$'' correspond to exterior and interior space-times, respectively. The extrinsic curvatures associated with the two surfaces of the shell region can be written as
\begin{equation}
K^{\pm}_{ij} =-n_\nu^{\pm}\left[\frac{\partial^2x_\nu}{\partial\xi^i\partial \xi^j}+\Gamma^\nu_{\alpha\beta} \frac{\partial x^\alpha}{\partial \xi^i}\frac{\partial x^\beta}{\partial \xi^j}\right], \label{eq:extrinsic}
\end{equation}
where the unit normal vector $n_\nu^{\pm}$ is defined as
\begin{equation}
n_\nu^{\pm}=\pm \bigg| g^{\alpha\beta}\frac{\partial f}{\partial x^\alpha}\frac{\partial f}{\partial x^\beta}\bigg|^{-1/2}\frac{\partial f}{\partial x^\nu}, \label{eq:normalvector}
\end{equation}
with $n^\nu n_\nu=1$ and $\xi^i$ is the instrinsic coordinate on the shell. $f$ is the metric component on the junction surface where we consider the metric simply as
\begin{equation}
ds^2 = -f(r)dt^2+f^{-1}(r)dr^2 +r^2(d\theta^2+\sin^2\theta d\phi^2).\label{eq:metriconjunction}
\end{equation}
Employing Lancozs equation (\ref{eq:Lancozs}), we can have the stress-energy surface tensor as $S^i_j =\text{diag}(-\sigma,q,q)$ at the junction $r=R$ where $\sigma$ and $q$ are surface energy density and surface pressure, respectively, where
\begin{eqnarray}
\sigma &=& - S^t_t = \frac{1}{\kappa}\left(K^t_t-K\right)=-\frac{2}{\kappa}K^\theta_\theta,\\
q &=& S^\theta_\theta = -\frac{1}{\kappa}\left(K^\theta_\theta-K\right)=\frac{1}{\kappa}\left(K^t_t +K^\theta_\theta\right).
\end{eqnarray}

Firstly, we consider the graviton condensate star type 1. The unit normal vectors to $r=R$ are
\begin{equation}
n_\nu^- = (0, \left(1-b-AR^2\right)^{-1/2},0,0), ~n_\nu^+ = (0, \left(1-\frac{2M}{R}\right)^{-1/2},0,0).\label{eq:normaltype1}
\end{equation}
By using (\ref{eq:normaltype1}), we can obtain
\begin{eqnarray}
\sigma_1 &=& -\frac{2}{\kappa R}\left(\sqrt{f}|^+-\sqrt{f}|^-\right)=0, \\ \label{eq:sigmatype1}
q_1 &=& -\frac{\sigma}{2}+\frac{1}{\kappa}\left(\frac{f'}{\sqrt{f}}\bigg|^+-\frac{f'}{\sqrt{f}}\bigg|^-\right)\nonumber\\
&=&\frac{1}{\kappa}\left[\frac{M}{R^2 \sqrt{1-\frac{2M}{R}}}-\frac{AR}{2\sqrt{1-b-AR^2}}\right].\label{qtype2}
\end{eqnarray}
We find that the surface energy density to vanish which means that the metric is continuous on the surface, so there is no mass contribution from the thin shell to the total mass $M$ of the graviton condensate star type 1. In addition, the normal derivative of the metric is discontinuous which corresponds to the presence of surface pressure $q_1$. This surface pressure $q_1$ compensates for the pressure jump and balance the pressure discontinuity on the surface of the graviton condensate star type 1.

Secondly, we consider the junction conditions for the graviton condensate type 2. This star takes simpler condition since it is clear that the metric continuous on the surface, implying that $\sigma_2=0$. Furthermore, the radial pressure is also continuous on the surface, implying that $q_2=0$ or the extrinsic curvature is still continuous. Hence, the existence of the thin shell is not required for type 2 because transverse pressure does not push on the boundary. It acts along directions tangent to the surface only ($T_{\theta\theta}n^\theta n^\theta =0$). Moreover, the transverse pressure also does not involve radial derivatives of the metric.

\section{Buchdahl Inequality}
\label{sec:Buchdahl}
Total energy density and total radial pressure in both stellar configurations possess a singular point at the center due to the existence of graviton condensate. In addition, there is another singularity on the pressures that can be calculated from the denominator of the pressures' equation. For type 1 of the graviton condensate star, the singularity occurs when $ 3\sqrt{1-b-AR^2}=\sqrt{1-b-AR_0^2} $ where $ r=R_0 $. This irregularity can occur when $ R_0 $ is given by
\begin{equation}
	R_0 = 3R \sqrt{1- \frac{8(1-b)R}{9(2m-bR)}}. \label{eq:irregularpoint}
\end{equation}
In order to obtain a real $ R_0 $, we need $ \frac{8(1-b)R}{9(2m-bR)}\leq 1 $. From this point, we may find the Buchdahl inequality as
\begin{equation}
	C <\frac{8+b}{18}. \label{eq:Buchdahl}
\end{equation}
It is obviously seen that the Buchdahl inequality depend on $ b $ for the stars. The upper limit of the inequality for type 1 is higher compared to that of CDS. Since it is assumed that $0\leq b\leq 1$, we can find the maximum compactness in the range $4/9 \leq C_{max} \leq 4.5/9$ for graviton condensate star type 1. The largest maximum compactness of this type is similar to the compactness of Schwarzschild BH while the smallest maximum compactness is similar to that of CDS. The smaller $b$ gives smaller $C_{max}$. It is worthwhile to note that the Buchdahl inequality is unaffected by the existence of the thin shell at the boundary, yet it is controlled entirely by the interior pressure profile.

For graviton condensate star type 2, the singularity occurs when $(3 +\frac{b}{AR^2})\sqrt{1-b-AR^2}=(1 +\frac{b}{AR^2})\sqrt{1-b-AR_0^2}$. We may employ the similar analysis to find the Buchdahl inequality where the following expression is generated,
\begin{equation}
	C <\frac{2+b+\sqrt{b^2-5b+4}}{9}. \label{eq:Buchdahl2}
\end{equation}
For graviton condensate star type 2, the upper limit is below the upper limit of CDS. When $b=0$, the maximum compactness $C_{max}$ is $4/9$ while when $b=1$ we can obtain that $C_{max}=3/9$. So, the range of the maximum compactness for all possible value of $b$ is $3/9 \leq C_{max} \leq 4/9$ for type 2. However, it implies that smaller $b$ gives larger $C_{max}$ in contrast with graviton condensate star type 1. The smallest $C_{max}$ of graviton condensate type 2 is similar to the compactness of Schwarzschild BH's photon sphere \citep{UrbanoVeermaeJCAP2019}.

\section{Gravitational Echoes}

It is believed that the gravitational waves' echoes exist on the stars within the ultra-compact regime. For CDS, this implies that the object's compactness must be in the range of $ 1/3\leq C \leq 4/9 $ \citep{MannarelliPRD2018,UrbanoVeermaeJCAP2019}. This range is obtained between the photon sphere and the Buchdahl limit. CDS is the simplest toy model to study the echo frequency and echo time of the gravitational echoes as it has been calculated in \cite{UrbanoVeermaeJCAP2019,PaniFerrariCQG2018}. The frequency of ultra-compact object is the highest compared to other fluid stars with the same mass and radius. The study of the gravitational echo time and frequency for a star containing dark energy have been examined in \cite{SaktiSulaksonoPRD2021}.

For graviton condensate star type 2, the procedure to investigate the gravitational echoes is similar to CDS because there is no thin shell. However, for graviton condensate star type 1, the existence of thin shell may need more investigations. We thank to \cite{PaniBertiPRD2099} who derive the gravitational perturbations for a stellar configuration especially and can be adapted in general for horizonless space-times and to static wormholes which includes a thin shell (see Appendix A in the reference for the detail derivation). For axial perturbations, the jump on the gravitational perturbation vanishes (Eq. (A35) in the reference) when metric components $g_{tt},g_{rr}$ are continuous between the exterior and interior space-times. So, the gravitational perturbations are continuous across the thin shell. This implies the continuity of the wave function in Regge-Wheeler equation and its derivative across the shell. Furthermore, since the graviton condensate star type 1 obeys the continuity on those metric components and we only consider axial gravitational perturbations in this work, the case will be similar to that of \cite{PaniBertiPRD2099}. Hence, the existence of the thin shell does not affect our computation on the axial perturbations in investigating the gravitational echoes.

We will investigate the gravitational echoes of the graviton condensate stars from its echo time, echo frequency, and the gravitational wave echoes in axial perturbation. In order to do so, we must work within the ultra-compact regime. From Eq. (\ref{eq:Buchdahl}), it is seen that the ultra-compact regime for graviton condensate star type 1 is given by
\begin{equation}
	\frac{1}{3}\leq C\leq \frac{8+b}{18} . \label{eq:ultracompact1}\
\end{equation}
Recall that our exterior solution is the Schwarzschild BH, hence the lower bound of this regime is $1/3$. For $b=1$, we have $1/3\leq C\leq 1/2$. For the graviton condensate star type 2, we then have
\begin{equation}
	\frac{1}{3}\leq C\leq \frac{2+b+\sqrt{b^2-5b+4}}{9}. \label{eq:ultracompact2}\
\end{equation}
For $b=1$, we have only one (ultra-)compactness $C=1/3$.

We now proceed to the calculation of the echo frequency and echo time. The echo frequency can be roughly estimated from the inverse of the echo time for a massless test scalar particle going from the photon sphere to the center of the ultra-compact object. The echo frequency is defined as $ f_{echo} \equiv \pi/\tau_{echo} $ while the echo time is calculated from \cite{UrbanoVeermaeJCAP2019,PaniFerrariCQG2018,MannarelliPRD2018}
\begin{equation}
	\tau_{echo} = \int^{3M}_{0} \left(-\frac{g_{rr}(r)}{g_{tt}(r)} \right)^{1/2} dr. \label{eq:tauformula}
\end{equation}
This echo time is computed from the center $r=0$ of the star to the surface at $r=R$ using the interior metric and from the surface to the unstable light ring using the exterior solution. For CDS, echo time and echo frequency have been computed in \cite{UrbanoVeermaeJCAP2019}. Using Eq. (\ref{eq:tauformula}) for graviton condensate star type 1, we obtain that
\begin{eqnarray}
	\frac{\tau_{echo1}}{M} &=& \frac{\cot^{-1}\left(\sqrt{\frac{\frac{4}{C} +\frac{b}{2C}-9}{1-\frac{b}{2C}}} \right) +\tan^{-1}\left(3/\sqrt{\frac{\frac{4}{C} +\frac{b}{2C}-9}{1-\frac{b}{2C}}}  \right) }{C^2\sqrt{\left(\frac{4}{C} +\frac{b}{2C}-9\right)\left(1-\frac{b}{2C}\right)}}\nonumber\\
	&-& 2\ln \left(\frac{1}{C}-2 \right)+3- \frac{1}{C}.\
\end{eqnarray}
Similarly for type 2, we obtain that
\begin{eqnarray}
	\frac{\tau_{echo2}}{M} &=&\frac{\cot^{-1}\left(\sqrt{\frac{4}{C} -9+b\left(\frac{2}{C}-\frac{1}{C^2}\right)} \right) }{C^2\sqrt{\frac{4}{C} -9+b\left(\frac{2}{C}-\frac{1}{C^2}\right)}}\nonumber\\
	&+&\frac{\tan^{-1}\left(\frac{3-b/C}{\sqrt{\frac{4}{C} -9+b\left(\frac{2}{C}-\frac{1}{C^2}\right)}} \right) }{C^2\sqrt{\frac{4}{C} -9+b\left(\frac{2}{C}-\frac{1}{C^2}\right)}}\nonumber\\
	&-& 2\ln \left(\frac{1}{C}-2 \right)+3- \frac{1}{C}.
\end{eqnarray}
Both echo times will be irregular when the compactness $ C=C_{irr}=1/2 $ which is the compactness of Schwazschild BH and at the Buchdahl limit. In addition, for graviton condensate star type 1, the irregularity can also occur when $ C=C_{irr}=b/2 $. In addition, for type 2, the irregularity also can occur when $ C=C_{irr}=\frac{2+b- \sqrt{b^2-5b+4}}{9} $, nonetheless, it is below the ultra-compact regime.

\begin{figure*}
	\centering
	\begin{tabular}{c c}
		\includegraphics[width=.45\linewidth]{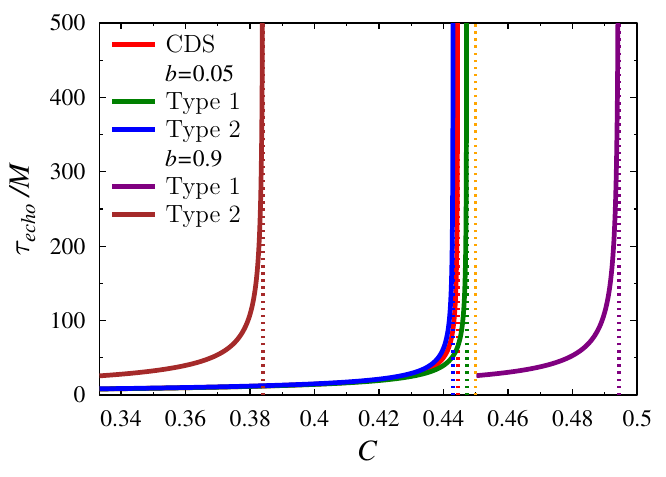} & \includegraphics[width=.45\linewidth]{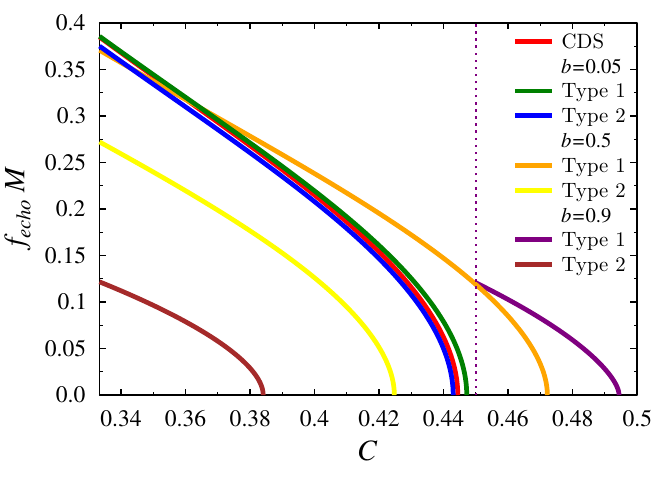}\\
		(a) & (b) \
	\end{tabular}
	\caption{(a) Plot of echo time as a function of $ C $ with $b=0$ (red) for CDS, $b=0.05$ (green for type 1, blue for type 2), and with $b=0.9$ (purple for type 1, brown for type 2). The dotted vertical lines denote the Buchdahl limit for $\tau_{echo}$, except for dotted orange line which denotes $C=b/2$. (b) Plot of echo frequency as a function of $ C $ with $b=0$ (red) for CDS, $b=0.1$ (orange for type 1, yellow for type 2), $b=0.5$ (purple for type 1, brown for type 2), and with $b=0.9$ (green for type 1, blue for type 2). The vertical dotted line denotes $C=b/2$.)}
	\label{fig:TauFreq}
\end{figure*}

To see the role of the parameter $ b $ in the gravitational echo's production, we compare the echo times and frequencies for both graviton condensate stars with CDS in Fig. \ref{fig:TauFreq}. Firstly, the echo time plots are shown in Fig. \ref{fig:TauFreq} (a). The echo time of CDS is given by the red line.  When $b=0.05$, at the same compactness, we find that $\tau_{echo2}$ (blue) $>$ $\tau_{CDS}$ (red) $>$ $\tau_{echo1}$ (green) and they will reach their own Buchdahl limit denoted by dotted vertical lines. For example at $C=0.4415$, we have $\tau_{echo2}/M=89.57$ $>$ $\tau_{CDS}/M=62.44$ $>$ $\tau_{echo1}/M=44.94$. When $b=0.9$, we can observe that  $\tau_{echo1}$ exists starting from $C>b/2=0.45$ because $C_{irr}=b/2$ is the irregular point. Hence, $\tau_{echo1}$ is positively and fully defined in the ultra-compact regime when $0\leq b\leq2/3$. So, in this range of $b$,  $\tau_{echo2}$ $>$ $\tau_{CDS}$ $>$ $\tau_{echo1}$ is always satisfied at the same compactness. It implies that the graviton condensate delays the echo production for type 2 and expedites the echo production for type 1 compared to CDS. Secondly, Fig. \ref{fig:TauFreq}  (b) shows the plots of echo frequencies for the stars as a function of compactness. $f_{echo}=0$ denotes the Buchdahl limit while there is another bound for graviton condensate star type 1 from $C_{irr}=b/2$. Since echo frequency is proportional to $1/\tau_{echo}$, it is inferred that $f_{echo2}$ $<$ $f_{CDS}$ $<$ $f_{echo1}$ when $0\leq b\leq2/3$ at the same compactness.

We have computed the echo time which denotes the time required for a massless test particle to travel from light ring to the center of the stars. This massless test particle travels in a potential well. This (effective) potential well denotes the existence of a second light ring. The simplest example of the effective potential of the ultra-compact object for CDS can be found in \cite{UrbanoVeermaeJCAP2019} where the minimum point of the potential is the stable light ring while the unstable one is the peak of the potential. To compute the effective potential, we may focus on the axial perturbation of the massless scalar field in the background of graviton condensate stars and their exterior solution (Schwarzschild BH). The perturbations of the scalar field are described by the following wave equation,
\begin{equation}
	\left[\frac{\partial^2}{\partial t^2} -\frac{\partial^2}{\partial r_*^2} +V_{s,l} (r)\right] \Psi_{s,l}(r_*,t)=0,\label{eq:waveequation}
\end{equation}
where $r_*$ is the tortoise coordinate given as $dr_* =e^{(\lambda-\nu)}dr$. The tortoise coordinate, in fact, measures the coordinate time that elapses along the radial null geodesics. $l$ is the azimuthal quantum number which complies $l\geq s$ where $s=0, \pm 1, \pm 2 $ for scalar, vector and tensor modes, respectively. The effective potentials for gravitational perturbation for the stars and its exterior solution (Schwarzschild BH) are given as follows \cite{KojimaYoshidaPTP1991,BoonsermNgampitipanPRD2013},
\begin{eqnarray}
&&V^{Int}_{2,l}(r)=e^{2\nu(r)}\left[\frac{l(l+1)}{r^2}-\frac{6M(r)}{r^3}+4\pi(\rho_{total}-p_r)\right],\label{eq:VeffGrav}\nonumber\\
\\
&&	V^{Ext}_{2,l}(r)=\frac{1}{r^3}\left(1-\frac{2M}{r}\right)\left[l(l+1)r-6M\right],\label{eq:VeffSch}
\end{eqnarray}
respectively. 

\begin{figure}
	\centering
	\includegraphics[width=.9\linewidth]{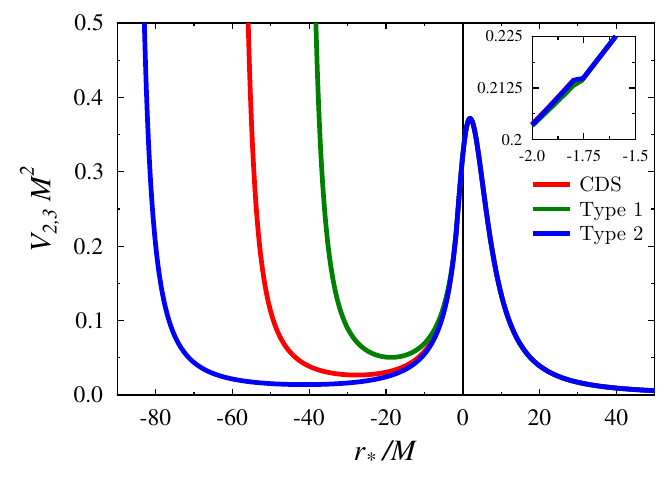} \
	\caption{Plot of effective potential for $s,l=2,3$ for CDS (red), graviton condensate stars type 1 (green), and type 2 (blue) for $b=0.05, C=0.4415$. The discontinuity (crack on the potential) denotes the star's surface.}
	\label{fig:Veff}
\end{figure}

We provide the plot of the effective potential in terms of $r_*$ in Fig. \ref{fig:Veff} where we choose $s,l=2,3$, $b=0.05$, and $C=0.4415$. Note that for $b=0.05$, the Buchdahl limit for graviton condensate stars type 1 is 0.4472 and type 2 is 0.443. Hence, $C=0.4415$ is still below the Buchdahl limit for all stars. We have exhibited that $\tau_{echo2} > \tau_{CDS} > \tau_{echo1}$ that can be proved also from the effective potential plots. From Fig. \ref{fig:Veff}, we can see that the width of the effective potential for type 1 is narrower than to that of CDS while the width of the effective potential for CDS is narrower than to that of type 2. The narrower effective potential, the lesser time needed for the scalar wave to travel within the effective potential. The discontinuity or crack on the effective potential denotes the position of the star's surface.

\begin{figure}
	\centering
	\includegraphics[width=1\linewidth]{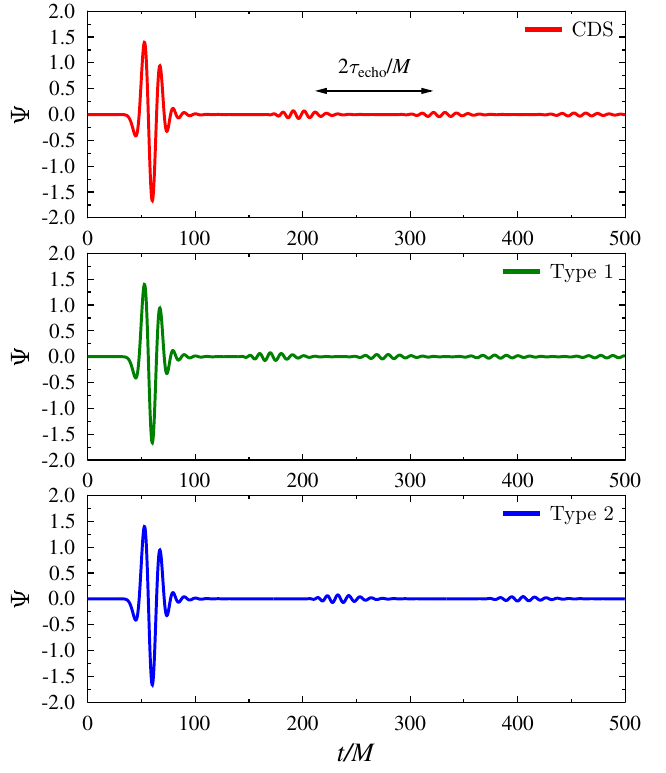} \
	\caption{Plot of wave form for $s,l=2,3$ for CDS (red), graviton condensate stars type 1 (green), and type 2 (blue) for $b=0.05, C=0.4415$. Time shown between two consecutive echoes is $2\tau_{echo}/M$.}
	\label{fig:Waveform}
\end{figure}

We also give the plot of the scalar wave form as a function of time in Fig. \ref{fig:Waveform} by solving Eq. (\ref{eq:waveequation}). To solve it, we need the following initial and boundary conditions:
\begin{eqnarray}
	&&\Psi (r_*,0)=0, ~~~~~~~~~~~~ \frac{\partial \Psi}{\partial t} (\infty,t)=-\frac{\partial \Psi}{\partial t} (\infty,t), \nonumber\\
	&&\frac{\partial \Psi}{\partial t} (r_*,0)=f(r_*), ~~~~~ \Psi (r_*^c,t)=0.\
\end{eqnarray}
The initial conditions on the left side describe the physical conditions of a post-merger phase with an initial Gaussian pulse centered at $r_*=r_g$ with spread $\sigma$ where $f(r_*)=exp\left(-(r_*-r_g)^2/\sigma^2\right)$ \citep{UrbanoVeermaeJCAP2019,JayawigunaBurikhamEPJC2024}. In our case, we choose $r_g=9M, \sigma=6M$. The boundary condition of the derivative of the scalar wave is to impose that only outgoing waves remain at spatial infinity while other boundary condition is for the regularity at the center of the star ($r^c_*\equiv r_*(r=0)$). The  time between two consecutive echoes in the figure is given as $2\tau_{echo}/M$. It is exhibited that the time between two consecutive echoes for CDS is larger than to that of graviton condensate star type 1 but smaller than to that of graviton condensate star type 2. Based on the calculation shown in Fig. \ref{fig:TauFreq}, we can obtain that $2\tau_{echo2}/M=179.14$ $>$ $2\tau_{CDS}/M=124.88$ $>$ $2\tau_{echo1}/M=89.88$ on Fig. {\ref{fig:Waveform}}.

\section{Summary}
This work has introduced novel exact interior solutions containing a perfect-fluid matter and a condensate of gravitons. To obtain the exact graviton condensate star solutions, we have assumed two distinct pressure conditions on the surface where type 1 is for $ p_{t}(R)=0 $ and type 2 is for $ p_{r}(R)=0 $. For graviton condensate star type 1, we have found that the thin shell needs to exist to balance the pressure's discontinuity while the type 2 does not need the thin shell. From the analysis of pressures' irregularity, we have found that the Buchdahl inequalities of these stars depend on $ b $. The Buchdahl inequality for graviton condensate type 1 is unaffected by the existence of the thin shell at the boundary, nevertheless it is controlled entirely by the interior pressure profile.

We then have considered the ultra-compact regime for which the compactnesses are in the range $1/3 \leq C \leq (8+b)/18  $ for type 1 and in the range $1/3 \leq C \leq (2+b+\sqrt{b^2-5b+4})/9 $ for type 2. So, the upper Buchdahl bound can be larger or smaller than to that of CDS depending on $b$. Since the graviton condensate stars could be an ultra-compact object in the arbitrary range of $C$ which depends on $b$, we have computed the time and frequency of the gravitational echoes. We have emphasized in the calculation that for type 1, the gravitational perturbation on the axial part does not alter by the existence of thin shell because the perturbation is continuous.

We have obtained that the presence of the graviton condensate could delay the gravitational echoes production for type 2 and could expedite it for type 1, compared to that of CDS. Our model of graviton condensate stars can be a fruitful toy model to study graviton condensate in the compact stellar objects. Moreover, since the future gravitational waves' detector will be available soon, the analysis of graviton condensate on the compact stellar objects will be possibly a new way to probe gravitons in the astrophysical compact objects.

\section*{Acknowledgements}
This research is supported by the Second Century Fund (C2F) and C2F research abroad scholarship, Chulalongkorn University, Thailand.






\end{document}